\documentstyle[12pt,epsfig]{article}
\def\bge{\begin{equation}}
\def\ene{\end{equation}}
\def\bg{\begin{eqnarray}}
\def\en{\end{eqnarray}}
\def\nn{\nonumber}
\def\ra{\rightarrow}
\def\la{\leftarrow}
\def\ua{\uparrow}
\def\da{\downarrow}
\def\del{\partial}
\def\qqbar{q \bar{q}}
\def\qbar{{\bar{q}}}
\def\ubar{{\bar{u}}}
\def\dbar{{\bar{d}}}
\def\sbar{{\bar{s}}}
\def\vk{\vec{k}}
\def\vr{\vec{r}}
\def\vx{\vec{x}}
\def\vy{\vec{y}}
\def\e{\epsilon}
\def\pomega{{\omega_B}}
\newcommand{\eq}{\label}
\newcommand{\non}{\nonumber}
\begin{document}
%%%%%%%%%%%%%%%%%%%%%%%%%%%%%%%%%%%%%%%%%%%%%%%%%%%%%%%%%%%%%%%%%%%%%%%%%%%%%
\begin{titlepage}
\title{Are $\eta$- and $\omega$-nuclear states bound~?}
\author{
K. Tsushima~\thanks{ktsushim@physics.adelaide.edu.au} ,
D.H. Lu~\thanks{dlu@physics.adelaide.edu.au} ,
A.W. Thomas~\thanks{athomas@physics.adelaide.edu.au}\\
{\small Special Research Center for the Subatomic Structure of Matter} \\
{\small and Department of Physics and Mathematical Physics} \\
{\small The University of Adelaide, SA 5005, Australia} \\
K. Saito~\thanks{ksaito@nucl.phys.tohoku.ac.jp} \\
{ \small Physics Division, Tohoku College of Pharmacy} \\
{\small Sendai 981-8558, Japan}  \\ }
\date{}
\maketitle
\vspace{-11cm}
\hfill ADP-98-28/T302
\vspace{11cm}
\begin{abstract}
We investigate theoretically whether it is feasible to detect $\eta$- and 
$\omega$-nucleus bound states. As well as the closed shell  
nuclei, $^{16}$O, $^{40}$Ca, $^{90}$Zr and $^{208}$Pb, 
we also investigate $^6$He, $^{11}$B and $^{26}$Mg, which are the final 
nuclei in the proposed experiment involving the (d,$^3$He) reaction at GSI.
Potentials for the $\eta$ and $\omega$ mesons in these nuclei
are calculated in local density approximation, embedding the mesons 
in the nucleus described by solving the mean-field equations of
motion in the QMC model. Our results suggest that one should expect
to find $\eta$- and $\omega$-nucleus bound states in all these nuclei.
\\ \\
{\it PACS}: 36.10.G, 14.40, 12.39.B, 21.65, 71.25\\
{\it Keywords}: Meson-nucleus bound states, Quark-meson coupling model,
In-medium meson properties, MIT bag model, Effective mass 
\end{abstract}
\end{titlepage}
%
%%%%%%%%%%%%%%%%%%%%%%%%%%%%%%%%%%%%%%%%%%%%%%%%%%%%%%%%%%%%%%%%%%%%%%%%%%%%%
%

The study of the properties of hadrons in a hot and/or dense nuclear   
medium is one of the most exciting new directions 
in nuclear physics. In particular, 
the medium modification of the light vector ($\rho$, $\omega$ and
$\phi$) meson masses has been investigated extensively by  
%many authors~\cite{qm97}--\cite{saitomega}.
many authors~\cite{qm97}. 
It has been suggested that dilepton production in the nuclear  
medium formed in relativistic heavy ion collisions, can provide a unique
tool to measure such modifications as meson mass shifts. 
For example, the experimental data  
obtained at the CERN/SPS by the CERES~\cite{ceres} and HELIOS~\cite{hel}
collaborations has been interpreted as evidence for a downward shift of the  
$\rho$ meson mass in dense nuclear matter~\cite{li}. 
To draw a more definite conclusion, 
measurements of the dilepton spectrum from vector
mesons produced in nuclei are planned at TJNAF~\cite{jlab} and
GSI~\cite{gsi}
% ,ins}.

Recently, a new, alternative approach to study meson mass shifts in nuclei
was suggested by Hayano {\it et al.}~\cite{hayano}.
Their suggestion is to use the (d, $^3$He) reaction to produce 
$\eta$ and $\omega$ mesons with nearly zero recoil.
If the meson feels a large enough, attractive (scalar) force inside
a nucleus, the meson is expected to form 
meson-nucleus bound states.  
Hayano {\it et al.}~\cite{hayano2} have estimated the binding energies
for various $\eta$-mesic nuclei. They have also 
calculated some quantities for the $\omega$ meson case. However, they   
used an $\eta$-nucleus optical potential calculated to first-order   
in density, taking as input the $\eta$-nucleon scattering length.  
In this article, we use an alternative, 
self-consistent method to study whether it is possible to form  
$\eta$- and $\omega$-nucleus bound states 
in $^{16}$O, $^{40}$Ca,  
$^{90}$Zr and $^{208}$Pb, as well as 
$^{6}$He, $^{11}$B and $^{26}$Mg. The latter three nuclei  
correspond to the proposed experiments at GSI~\cite{hayano}
using the (d,$^3$He) reaction -- i.e., 
the reactions, $^7$Li\,(d,$^3$He)\,$^6_{\eta/\omega}$He,  
$^{12}$C\,(d,$^3$He)\,$^{11}_{\eta/\omega}$B and
$^{27}$Al\,(d,$^3$He)\,$^{26}_{\eta/\omega}$Mg.

In earlier work
we addressed the question of whether quarks play 
an important role in finite nuclei~\cite{finite0,finite1,finite2}. 
The quark-meson coupling (QMC) model~\cite{gui,finite0}, 
which is based explicitly on quark 
degrees of freedom,  
is probably one of the most appropriate models to study whether 
meson-nucleus bound states are possible.
The model has been able to describe successfully the static properties 
of both nuclear matter and finite nuclei~\cite{finite1,appl}, 
as well as meson properties in the nuclear 
medium~\cite{finite2}. 
Thus, the model is ideally suited to treat a bound meson and the 
nucleons in a nucleus on the same footing.
In this study, we investigate the possible formation of the 
$\eta$- and $\omega$-nucleus bound states due to downward shifts of the 
masses. We will use QMC-I~\cite{finite2}, where the effective, 
isoscalar-vector $\omega$ field, which is off mass-shell and 
mediates the interactions among nucleons,   
is distinguished from the physical (on mass-shell) 
$\omega$ meson which is produced inside the nuclei by the above
mentioned experiments.

One of the most attractive features of QMC is 
that, in practice, it is not significantly more complicated than 
Quantum Hadrodynamics (QHD)~\cite{qhd}, although the quark 
substructure of hadrons is explicitly implemented. 
Furthermore, it produces a reasonable value for 
the nuclear incompressibility.
A detailed description of the Lagrangian density, and the 
mean-field equations of motion needed to describe a finite nucleus,
%is given in Refs.~\cite{finite0,finite1,hyper}.
is given in Refs.~\cite{finite0,finite1,finite2}.

At position $\vr$ in a nucleus (the coordinate origin is taken at 
the center of the nucleus), 
the Dirac equations for the quarks and antiquarks 
in the $\eta$ and $\omega$ meson bags are given 
by~\cite{finite2}:
\bge
\left[ i \gamma \cdot \partial_x - (m_q - V_\sigma(\vr)) \mp \gamma^0
\left( V_\omega(\vr) + \frac{1}{2} V_\rho(\vr) \right) \right]
\left(\begin{array}{c} \psi_u(x)\\ \psi_\ubar(x)\\ \end{array}\right) 
 = 0,
\label{diracu}
\ene
\bge
\left[ i \gamma \cdot \partial_x - (m_q - V_\sigma(\vr)) \mp \gamma^0
\left( V_\omega(\vr) - \frac{1}{2} V_\rho(\vr) \right) \right]
\left(\begin{array}{c} \psi_d(x)\\ \psi_\dbar(x)\\ \end{array} \right) 
 = 0,
\label{diracd}
\ene
\bge
\left[ i \gamma \cdot \partial_x - m_s \right] 
\psi_s (x)\,\, ({\rm or}\,\, \psi_\sbar(x)) = 0.
\label{diracs}
\ene
(Note that we have neglected a possible, very slight variation of the
scalar and vector mean-fields inside the meson bag 
due to its finite size~\cite{finite0}.)  
The mean-field potentials for a bag centered 
at position $\vr$ in the nucleus, which will be calculated 
self-consistently, are defined by,
$V_\sigma(\vr) = g^q_\sigma \sigma(\vr), 
V_\omega(\vr) = g^q_\omega \omega(\vr)$ and
$V_\rho(\vr) = g^q_\rho b(\vr)$, with $g^q_\sigma, g^q_\omega$ 
and $g^q_\rho$ being, respectively, the corresponding  
quark and meson-field coupling constants.
Here we assume 
that the current masses
are given as 
$m_q \equiv m_u = m_d = m_{\ubar} = m_{\dbar}$.
Furthermore, we have assumed that the $\sigma$, $\omega$ and $\rho$ 
fields only interact directly with the nonstrange quarks and 
antiquarks~\cite{finite2}. 
The mean meson fields at position $\vr$ in the nucleus   
are calculated self-consistently by solving Eqs.~(23) -- (30) of 
Ref.~\cite{finite1}.

Hereafter we use the notation, $\pomega$, to specify the physical,  
bound $\omega$ meson, in order to avoid confusion
with the isoscalar-vector $\omega$ field appearing in QMC.
The static solution for the ground state quarks or 
antiquarks in the $\eta$ and $\omega$ meson bags may be written as: 
\bge
\psi_f (x) = N_f e^{- i \epsilon_f t / R_j^*} \psi_f (\vx),
\qquad ({\rm for}\;\; j = \eta, \pomega \;\; {\rm and}\;\; 
f = u, \ubar, d, \dbar, s, \sbar), 
\label{wavefunction}
\ene
where $N_f$ and $\psi_f (\vx)$ are respectively the normalization factor
and corresponding spin and spatial part of the wave function~\cite{finite2}.
The bag radius in medium, $R_j^* \,\,(j=\eta,\pomega)$, 
which depends on the hadron species in which the quarks 
and antiquarks belong, will be determined self-consistently
through the stability condition for the (in-medium) mass of the meson 
%against the variation of the bag radius~\cite{finite1,finite2,hyper}.
against the variation of the bag radius.
(See Eq.~(\ref{equil}) below.)
The eigenenergies for the quarks, in units of $1/R_j^*$, 
%$\epsilon_f\;
%(f=u,\ubar,d,\dbar,s,\sbar)$ in Eq. (\ref{wavefunction}), 
are given by
\bg
\left( \begin{array}{c} \e_u(\vr) \\ \e_{\ubar}(\vr) \end{array} \right)
&=& \Omega_q^*(\vr) \pm R_j^* \left(
V_\omega(\vr) + \frac{1}{2} V_\rho(\vr) \right), \nn 
\\
\left( \begin{array}{c} \e_d(\vr) \\ \e_{\dbar}(\vr) \end{array} \right)
&=& \Omega_q^*(\vr) \pm R_j^* \left(
V_\omega(\vr) - \frac{1}{2} V_\rho(\vr) \right), \nn
\\
\e_s(\vr) &=& \e_{\sbar}(\vr) = \Omega_s(\vr),
\label{quarkenergy}
\en
%
%where $\Omega_q^*(\vr) = \sqrt{x_q^2 + (R_j^* m_q^*)^2}$ with
%$m_q^* = m_q - g^q_\sigma \sigma(\vr)\, (q = u, \ubar, d, \dbar)$ and 
%$\Omega_s(\vr) = \sqrt{x_s^2 + (R_j^* m_s)^2}$. 
where $\Omega_q^*(\vr) = \sqrt{x_q^2 + (R_j^* m_q^*)^2}$ 
and $\Omega_s(\vr) = \sqrt{x_s^2 + (R_j^* m_s)^2}$ 
with $m_q^* = m_q - g^q_\sigma \sigma(\vr) (q = u, \ubar, d, \dbar)$. 
The bag eigenfrequencies, $x_q$ and $x_s$, are 
determined by the usual, linear boundary condition~\cite{finite0}. 

Next, we consider the $\eta$ and $\pomega$ meson masses in the nucleus.
The physical states of the $\eta$ and $\pomega$ mesons are the 
superpositions of the octet and singlet states:
% with the mixing angles, 
%$\theta_P$ and $\theta_V$ for the $\eta - \eta'$ and 
%$\phi - \pomega$ mixings, respectively~\cite{pdata}:
%
\bg
\xi  &=& \xi_8 \cos\theta_{P,V} - \xi_1 \sin\theta_{P,V},\quad
\xi' = \xi_8 \sin\theta_{P,V} + \xi_1 \cos\theta_{P,V}, 
\label{mixing1}\\
{\rm with}\hspace{2cm} \nn
\\
\xi_1 &=& \frac{1}{\sqrt{3}}\; (u\ubar + d\dbar + s\sbar),\quad
\xi_8 = \frac{1}{\sqrt{6}}\; (u\ubar + d\dbar - 2 s\sbar), 
\label{mixing2}
%\\
%\left( \begin{array}{c} \xi \\ \xi' \end{array} \right)
%&=&  \left( \begin{array}{c} \eta \\ \eta' \end{array} \right)
%\quad {\rm or}\quad
%\left( \begin{array}{c} \phi \\ \pomega \end{array} \right),
%\quad (\theta_P\; {\rm for}\; \eta-\eta'\;\;{\rm and}\;\;
%\theta_V\; {\rm for}\; \phi-\pomega),\qquad
%\label{mixing3}
\en
\noindent
where ($\xi$, $\xi'$) denotes ($\eta$, $\eta'$) or ($\phi$, $\pomega$),
with the mixing angles 
$\theta_P$ or $\theta_V$, respectively~\cite{pdata}.
%for the $\eta - \eta'$ and 
%$\theta_V$ for $\phi - \pomega$ mixings, respectively~\cite{pdata}:
Then, the masses for the $\eta$ and $\pomega$ mesons in  
the nucleus at the position $\vr$, are self-consistently calculated by:
\bg
m_\eta^*(\vr) &=& \frac{2 [a_P^2\Omega_q^*(\vr) 
+ b_P^2\Omega_s(\vr)] - z_\eta}{R_\eta^*}
+ {4\over 3}\pi R_\eta^{* 3} B,
\label{meta}\\
m_\pomega^*(\vr) &=& \frac{2 [a_V^2\Omega_q^*(\vr) 
+ b_V^2\Omega_s(\vr)] - z_\pomega}{R_\pomega^*}
+ {4\over 3}\pi R_\pomega^{* 3} B,
\label{momega}\\
\left.\frac{\partial m_j^*(\vr)}
{\partial R_j}\right|_{R_j = R_j^*} &=& 0, \quad\quad (j = \eta,\pomega), 
\label{equil}\\
{\rm with}\qquad \nn \\
a_{P,V} &=& \frac{1}{\sqrt{3}} \cos\theta_{P,V} 
- \sqrt{\frac{2}{3}} \sin\theta_{P,V},\quad
b_{P,V} = \sqrt{\frac{2}{3}} \cos\theta_{P,V}
+ \frac{1}{\sqrt{3}} \sin\theta_{P,V}. 
\label{abdeff}
\en
%
%For the values of the mixing angles, 
In practice, we use $\theta_P = - 10^\circ$ and 
$\theta_V = 39^\circ$~\cite{pdata}, neglecting any possible mass 
dependence and imaginary parts.
We also assume that the values of the mixing angles do not change in 
medium, although this is possible and merits further investigation.
In Eqs.~(\ref{meta}) and (\ref{momega}), $z_\eta$ and 
$z_\pomega$ parameterize the sum of the center-of-mass and gluon 
fluctuation effects, and are assumed to be independent of 
density~\cite{finite0}. 
%The parameters are determined in free space 
%using the physical states for the $\eta$ and $\pomega$ mesons 
%(i.e., using the extracted mixing angles, 
%$\theta_P = - 10^\circ$ and
%$\theta_V = 39^\circ$, respectively). 

In this study, we chose $m_q = 5$ MeV and $m_s = 250$ MeV,
for the current quark masses, and $R_N = 0.8$ fm
for the bag radius of the nucleon in free space.
Other inputs, parameters, and some of the quantities calculated
in the present study, are listed in Table~\ref{bagparam}.
%We stress that while the model has a number of parameters, 
%only three of them, 
The coupling constants, $g^q_\sigma$, $g^q_\omega$
and $g^q_\rho$, are adjusted to fit 
%nuclear data -- namely 
the saturation energy and density of symmetric nuclear matter and the 
bulk symmetry energy. Note that none of the results for nuclear properties 
depend strongly on the choice
of the other parameters -- for example, the
relatively weak dependence of the final results on the values
of the current quark mass and bag radius is shown explicitly in
Refs.~\cite{finite0,finite1}.
The parameters at the hadronic level associated with the core nucleus
are summarized in Table~\ref{hparamt}.
%Concerning the parameters for the $\sigma$ field, we note that
%the properties of nuclear matter only fix the ratio, ($g_\sigma/m_\sigma$),
%with a chosen value, $m_\sigma = 550$ MeV.
%Keeping this ratio to be a constant,
The value of the $\sigma$ mass for finite nuclei
is obtained by fitting the r.m.s. charge radius of $^{40}$Ca
to the experimental value, $r_{{\rm ch}}(^{40}$Ca) = 3.48 fm~\cite{finite1}.
For more details and explanations of the model parameters,
see Refs.~\cite{finite0,finite1}.
%
%%%%%%%%%%%%%%%%%%%%%%%%%%%%%%%%%%%%%%%%%%%%%%%%%%%%%%%%%%%%%%%%%%%%%%%
\begin{table}[htbp]
\begin{center}
\caption{
Physical masses fitted in free space, free space full widths, $\Gamma$, 
the bag parameters, $z$, and the bag radii in free space, $R$.
The quantities with an asterisk, are those quantities calculated
at normal nuclear matter density, $\rho_0 = 0.15$ fm$^{-3}$.
They are obtained with the bag constant, $B = (170$ MeV$)^4$, current 
quark masses, $m_u = m_d = 5$ MeV and $m_s = 250$ MeV. Note that the free
space width of the $\eta$ meson is 1.18 keV~\protect\cite{pdata}.}
\label{bagparam}
\begin{tabular}[t]{c|lcclll}
\hline
&mass (MeV) &$\Gamma$ (MeV) &$z$ &$R$ (fm)& $m^*$ (MeV) & $R^*$ (fm)\\ 
\hline 
$N$       &939.0 (input) & --- &3.295       &0.800 (input) &754.5 &0.786\\
$\eta$    &547.5 (input) &0 (input)  &3.131 &0.603 &483.9 &0.600\\
$\pomega$ &781.9 (input) &8.43 (input)      &1.866 &0.753 &658.7 &0.749\\
\hline
\end{tabular}
\end{center}
\end{table}
%
%%%%%%%%%%%%%%%%%%%%%%%%%%%%%%%%%%%%%%%%%%%%%%%%%%%%%%%%%%%%%%%%%%%%%%%
%hadronic level parameters
\begin{table}[htbp]
\begin{center}
\caption{Parameters at the hadronic
level (masses and coupling constants of mesons and photon 
for finite nuclei)~\protect\cite{finite1}.}
\label{hparamt}
\begin{tabular}[t]{c|cc}
\hline
field &mass (MeV) &$g^2/4\pi\, (e^2/4\pi)$\\
\hline
$\sigma$ &418 &3.12\\
$\omega$ &783 &5.31\\
$\rho$   &770 &6.93\\
$A$      &0   &1/137.036\\
\hline
\end{tabular}
\end{center}
\end{table}
%
%%%%%%%%%%%%%%%%%%%%%%%%%%%%%%%%%%%%%%%%%%%%%%%%%%%%%%%%%%%%%%%%%%%%%%
%

\begin{figure}[hbt]
\begin{center}
\epsfig{file=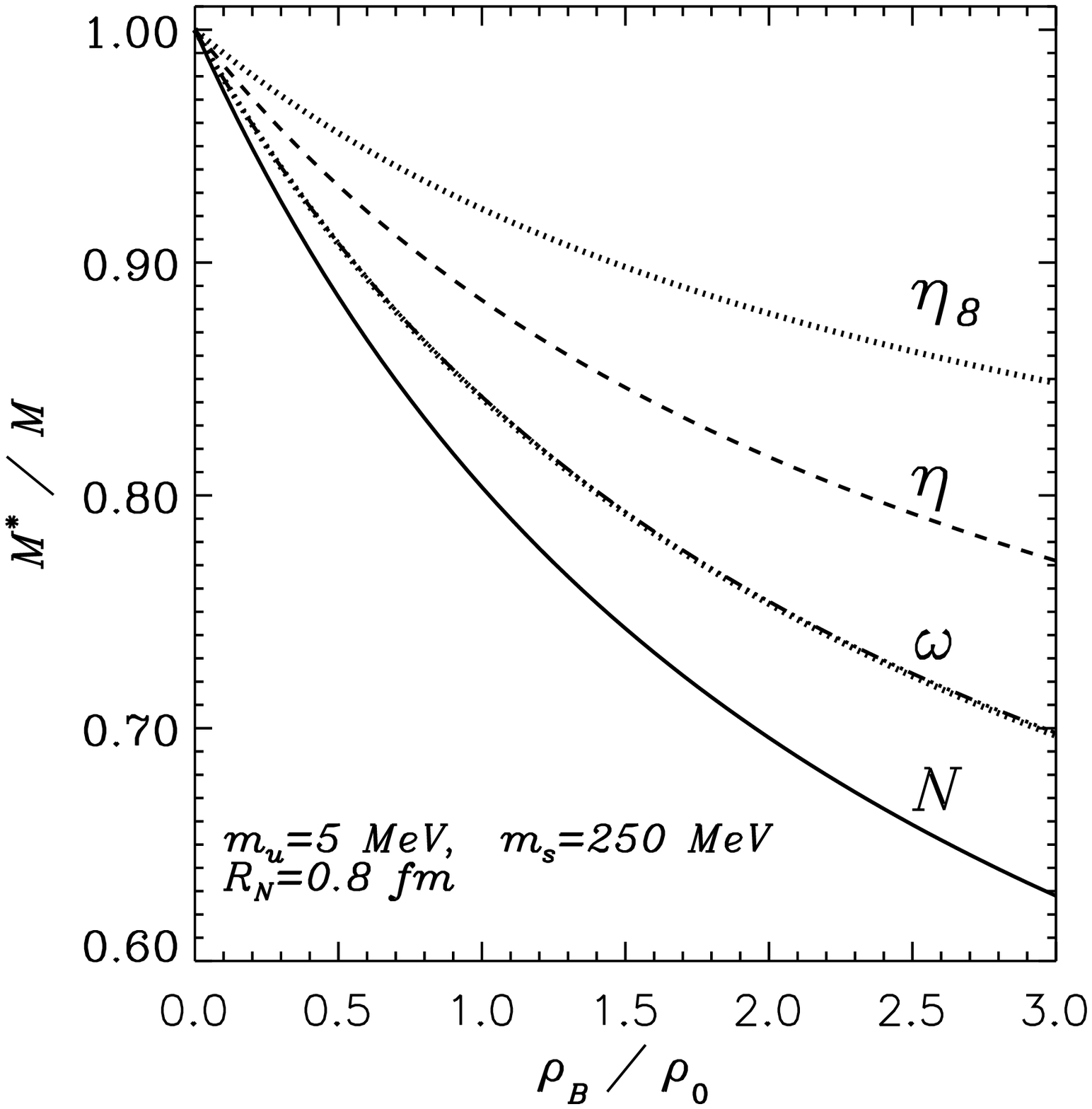,height=15cm}
\caption{Effective masses of the nucleon, physical $\eta$ and $\omega$ 
mesons, and those calculated based on SU(3) quark model basis 
(the dotted lines), namely
$\omega = \frac{1}{\protect\sqrt{2}} (u\ubar + d\dbar)$ (ideal mixing)  
and $\eta_8 = \frac{1}{\protect\sqrt{6}} (u\ubar + d\dbar - 2 s\sbar)$. 
The two cases for the $\omega$ meson are almost degenerate.
(Normal nuclear matter density, $\rho_0$, is 0.15 fm$^{-3}$.)}
\label{etaomass}
\end{center}
\end{figure}

Through Eqs.~(\ref{diracu}) -- (\ref{abdeff}) we self-consistently  
calculate effective masses, $m^*_\eta(\vr)$ and $m^*_\pomega(\vr)$  
at the position $\vr$ in the nucleus. 
Because the vector potentials for 
the same flavor of quark and antiquark cancel each other,  
the potentials for the $\eta$ and $\pomega$ mesons are given respectively by
$m^*_\eta(r) - m_\eta$ and  $m^*_\pomega(r) - m_\pomega$, 
where they will depend only on the distance from the center of the 
nucleus, $r = |\vr|$.
Before showing the calculated potentials for the $\eta$ and $\pomega$, 
we first show in Fig.~\ref{etaomass} their effective masses 
and those calculated within an SU(3) quark model basis,
$\omega = \frac{1}{\sqrt{2}} (u\ubar + d\dbar)$ (ideal mixing) and 
$\eta_8 = \xi_8$ in Eq.~(\ref{mixing2}), in symmetric nuclear matter.
One can easily see that the effect of the singlet-octet mixing 
is negligible for the $\pomega$ mass in matter,
whereas it is important for the $\eta$ mass. 

\begin{figure}[hbt]
\begin{center}
\hspace*{-0.8cm}
\epsfig{file=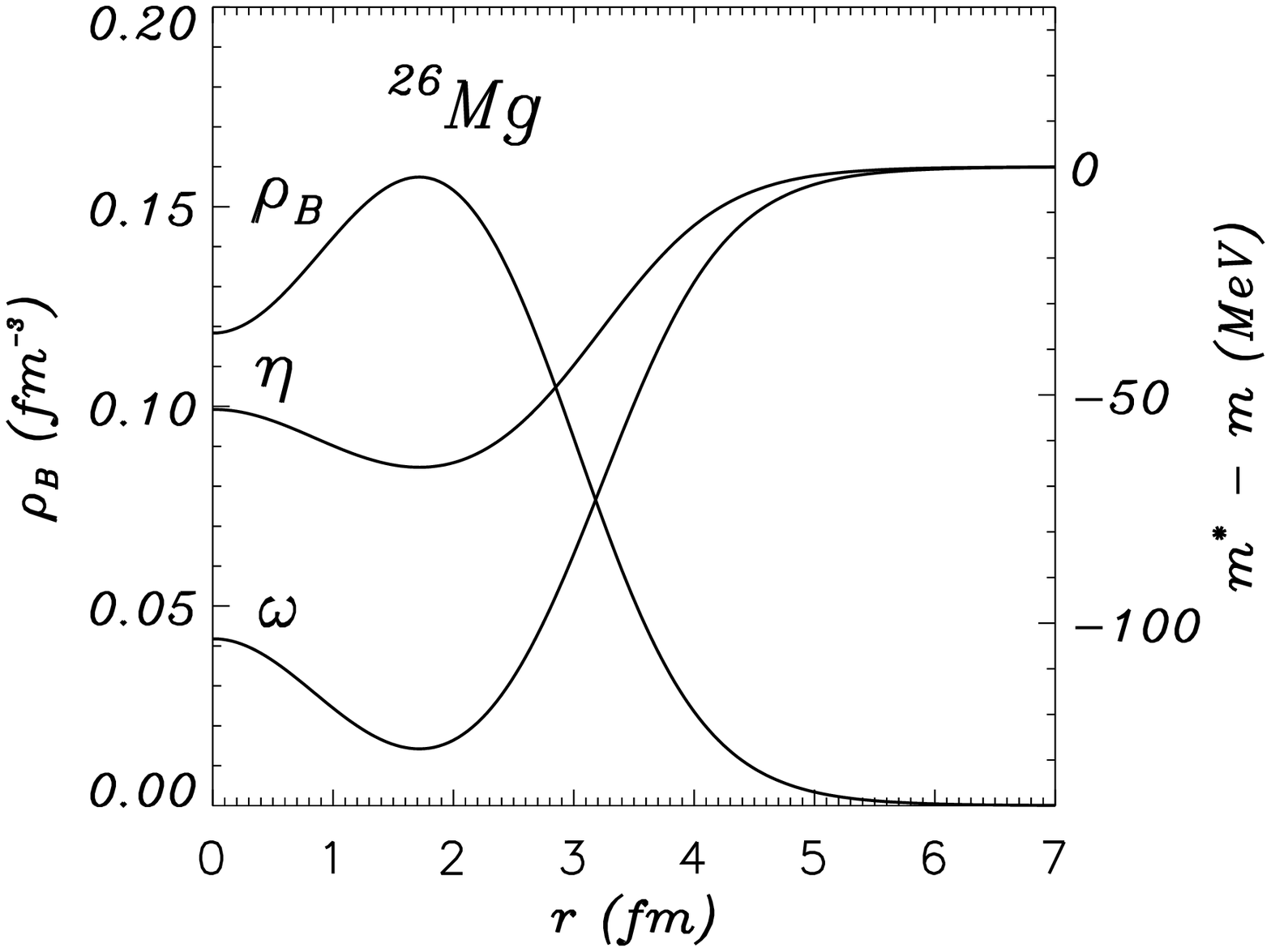,height=8cm}\,\quad 
\hspace*{-0.8cm}
\epsfig{file=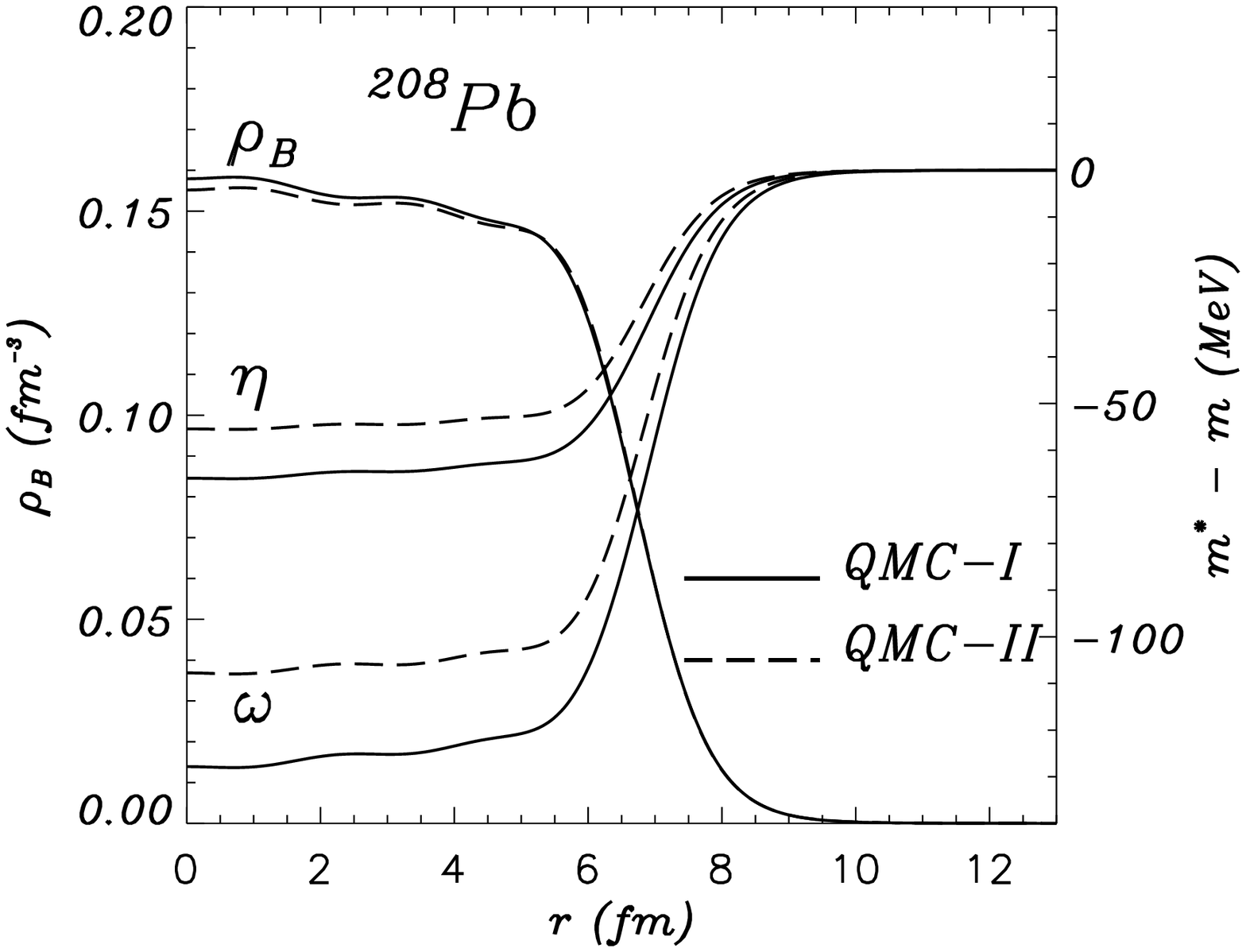,height=8cm}
\caption{Potentials for the $\eta$ and $\omega$ mesons,  
($m^*_\eta(r) - m_\eta$) and ($m^*_\pomega(r) - m_\pomega$), calculated 
in QMC-I for $^{26}$Mg and $^{208}$Pb.
%$^{16}$O, $^{40}$Ca $^{90}$Zr and $^{208}$Pb. 
For $^{208}$Pb 
the potentials are also shown for QMC-II.}
\label{ocazrpb}
\end{center}
\end{figure}

%Next we show the potentials for the mesons in various nuclei calculated
%self-consistently for $^{16}$O, $^{40}$Ca, $^{90}$Zr and
%$^{208}$Pb, and $^{6}$He, $^{11}$B and $^{26}$Mg,
%in Figs.~\ref{ocazrpb} and~\ref{hebmg}, respectively.
As an example, we show the potentials for the mesons in 
$^{26}$Mg and $^{208}$Pb in Fig.~\ref{ocazrpb}. 
Note that the actual calculations for $^{6}$He, $^{11}$B and $^{26}$Mg are
performed in the same way as for the closed shell nuclei,
$^{16}$O, $^{40}$Ca, $^{90}$Zr and $^{208}$Pb. 
Although $^{6}$He, $^{11}$B and $^{26}$Mg are not spherical, 
we have neglected the effect of deformation, which is expected to be
small and irrelevant for the present discussion. (We do not expect that 
deformation should alter the calculated potentials by more than 
a few MeV near the center of the deformed nucleus, because 
the baryon (scalar) density there is also expected 
to be more or less the same as that for a spherical nucleus -- 
close to normal nuclear matter density.)
The depth of the potentials are typically 60 and 130 MeV  
for the $\eta$ and $\pomega$ mesons, respectively, 
around the center of each nucleus.
In addition, we show the calculated potentials using QMC-II~\cite{finite2}  
in Fig.~\ref{ocazrpb}, for $^{208}$Pb, in order to estimate   
the ambiguities due to different versions of the QMC model.
At the center of $^{208}$Pb, the potential calculated using QMC-II is 
about 20 MeV shallower than that for QMC-I. 

Now we are in a position to calculate single-particle energies for the 
mesons using the potentials calculated in QMC.
Because the typical momentum of the bound $\omega$ is low, it should be 
a very good approximation to neglect the possible energy difference 
between the longitudinal and transverse components of the 
$\omega$~\cite{saitomega}.
Then, after imposing the Lorentz condition, $\partial_\mu \phi^\mu = 0$,  
solving the Proca equation becomes equivalent to solving the Klein-Gordon
equation,
% Thus, to obtain the meson nucleus binding energies, 
% we may solve the Klein-Gordon equations: 
%
\bge
\left[ \nabla^2 + E^2_j - m^{*2}_j(r) \right]\,
\phi_j(\vr) = 0, \qquad (j=\eta,\pomega),
\label{kgequation1}
\ene
where $E_j$ is the total energy of the meson.
An additional complication, which has so far been ignored, is the 
meson absorption in the nucleus, which requires a complex potential. 
%an imaginary part for the potential.
% to describe the effect. 
At the moment, we have not been able to calculate the imaginary part 
of the potential (equivalently, the in-medium widths of the mesons)
self-consistently within the model. 
In order to make a more realistic estimate for the meson-nucleus bound states,
we include the widths of the $\eta$ and $\pomega$ mesons in the nucleus 
by assuming a specific form: 
\bg
\tilde{m}^*_j(r) &=&
m^*_j(r) - \frac{i}{2} 
\left[ (m_j - m^*_j(r)) 
\gamma_j + \Gamma_j \right], \qquad (j=\eta,\pomega),
\label{imaginary}\\
&\equiv& m^*_j(r) - \frac{i}{2} \Gamma^*_j (r),
\label{width}
\en
where, $m_j$ and $\Gamma_j$ are the corresponding masses and widths 
in free space listed in Table~\ref{bagparam}, and
%In Eq.~(\ref{imaginary}) 
$\gamma_j$ are treated as phenomenological 
parameters to describe the in-medium meson widths,
$\Gamma^*_j(r)$. 
% $\equiv (m_j - m^*_j(r)) \gamma_j + \Gamma_j$.
According to the estimates in Refs.~\cite{hayano,fri}, 
the widths of the mesons in nuclei and at normal nuclear matter density 
are $\Gamma^*_\eta \sim 30 - 70$ MeV~\cite{hayano}
and $\Gamma^*_\pomega \sim 30 - 40$ MeV~\cite{fri}, respectively.
Thus, we calculate the single-particle energies for several values 
of the parameter, $\gamma_j$,
%s appearing in Eq.~(\ref{imaginary}), 
%$\gamma_\eta = 0, 0.5, 1.0$ 
%and $\gamma_\pomega = 0, 0.2, 0.4$ 
which cover the estimated ranges. 

{} From Table~\ref{bagparam} and the calculated density distributions 
%Figs.~\ref{ocazrpb} and~\ref{hebmg}, 
one can obtain the corresponding widths at normal nuclear matter 
density, as well as in the finite nuclei.
% and also in the corresponding nucleus which vary  
%in position, or density. 
Because of the recoilless condition for meson production in the 
GSI experiment~\cite{hayano,hayano2}, we may expect that the energy 
dependence of the potentials would not be strong~\cite{pion}.  
Thus we actually solve the following, modified Klein-Gordon equations:
\bge
\left[ 
\nabla^2 + E^2_j - \tilde{m}^{*2}_j(r) 
\right]\, \phi_j(\vr) = 0, \qquad (j=\eta,\pomega).
\label{kgequation2}
\ene
%
%Eq.~(\ref{kgequation2}) has been solved 
This is carried out in momentum 
space by the method developed in Ref.~\cite{landau}.
To confirm the calculated results, we also calculated the 
single-particle energies by solving 
the Schr\"{o}dinger equation.
Calculated single-particle energies for the $\eta$ and $\pomega$ 
mesons, obtained solving the Klein-Gordon equation  
are respectively listed in Tables~\ref{etaenergy} 
and~\ref{omegaenergy}. 
%(Note that the advantage of solving the Klein-Gordon 
%equation in momentum space is that it can handle quadratic terms
%arising in the potentials without any trouble, as 
%was demonstrated in Ref.~\cite{landau}.) 

%%%%%%%%%%%%%%%%%%%%%%%%%%%%%%%%%%%%%%%%%%%%%%%%%%%%%%%%%%%%%%%%%%%%%%
%\newpage
%Eta
\begin{table}[htbp]
\begin{center}
\caption{
Calculated $\eta$ meson single-particle energies, 
$E = Re(E_\eta - m_\eta)$, 
and full widths, $\Gamma$, (both in MeV), in various nuclei, where 
the complex eigenenergies are, $E_\eta = E + m_\eta - i \Gamma/2$. 
See Eq.~(\protect\ref{imaginary}) for 
the definition of $\gamma_\eta$. Note that the free space width 
of the $\eta$ is 1.18 keV, which corresponds 
to $\gamma_\eta = 0$.
}

\label{etaenergy}
\begin{tabular}[t]{lc|cc|cc|cc}
\hline \hline
& &$\gamma_\eta$=0 & &$\gamma_\eta$=0.5& &$\gamma_\eta$=1.0& \\
\hline \hline
& &$E$ &$\Gamma$ &$E$ &$\Gamma$ &$E$ &$\Gamma$ \\
\hline
$^{16}_\eta$O &1s &-33.1 &0 &-32.6&26.7 &-31.2&53.9 \\
              &1p &-8.69 &0 &-7.72&18.3 &-5.25&38.2 \\
\hline
$^{40}_\eta$Ca &1s &-46.5 &0 &-46.0&31.7 &-44.8&63.6 \\
               &1p &-27.4 &0 &-26.8&26.8 &-25.2&54.2 \\
               &2s &-6.09 &0 &-4.61&17.7 &-1.24&38.5 \\
\hline
%
%$^{48}_\eta$Ca &1s &-47.2 &0  &-46.8&31.4 &-45.7&62.9 \\
%               &1p &-30.3 &0  &-29.8&27.5 &-28.4&55.3 \\
%               &2s &-8.97 &0  &-7.78&19.8 &-4.93&41.9 \\
%\hline
%
$^{90}_\eta$Zr &1s &-53.3 &0 &-52.9&33.2 &-51.8&66.4 \\
               &1p &-40.5 &0 &-40.0&30.5 &-38.8&61.2 \\
               &2s &-22.3 &0 &-21.7&26.1 &-19.9&53.1 \\ 
\hline
$^{208}_\eta$Pb &1s &-56.6 &0 &-56.3&33.2 &-55.3&66.2 \\
                &1p &-48.7 &0 &-48.3&31.8 &-47.3&63.5 \\
                &2s &-36.3 &0 &-35.9&29.6 &-34.7&59.5 \\
\hline
\hline
$^{6}_\eta$He &1s &-11.4 &0 &-10.7&14.5 & -8.75&29.9 \\
\hline
$^{11}_\eta$B &1s &-25.0 &0 &-24.5&22.8 &-22.9&46.1 \\
\hline
$^{26}_\eta$Mg &1s &-39.2 &0 &-38.8&28.5 &-37.6&57.3 \\
               &1p &-18.5 &0 &-17.8&23.1 &-15.9&47.1 \\
\hline \hline
\end{tabular}
\end{center}
\end{table}
%
%%%%%%%%%%%%%%%%%%%%%%%%%%%%%%%%%%%%%%%%%%%%%%%%%%%%%%%%%%%%%%%%%%%%%%
%\newpage
%Omega
\begin{table}[htbp]
\begin{center}
\caption{ 
%Calculated $\omega$ meson single-particle energies, 
%$E = Re(E_\omega - m_\omega)$, 
%and full widths, $\Gamma$, (both in MeV) in various nuclei, where 
%the complex eigenenergies are, $E_\omega = E + m_\omega - i \Gamma/2$. 
%See Eq.~(\protect\ref{imaginary}) for 
%the definition of $\gamma_\omega$. 
As in Tables~\protect\ref{etaenergy}, but for $\omega$ meson 
single-particle energies.  
In the light of $\Gamma$ in 
Refs.~\protect\cite{fri}, the results with $\gamma_\omega = 0.2$ are
expected to correspond best with the experiment.
}
\label{omegaenergy}
\begin{tabular}[t]{lc|cc|cc|cc}
\hline \hline
& &$\gamma_\omega$=0 & &$\gamma_\omega$=0.2& &$\gamma_\omega$=0.4& \\
\hline \hline
& &$E$ &$\Gamma$ &$E$ &$\Gamma$ &$E$ &$\Gamma$ \\
\hline
$^{16}_\omega$O &1s &-93.5&8.14 &-93.4&30.6 &-93.4&53.1 \\
                &1p &-64.8&7.94 &-64.7&27.8 &-64.6&47.7 \\
\hline
$^{40}_\omega$Ca &1s &-111&8.22  &-111&33.1  &-111&58.1 \\
                 &1p &-90.8&8.07 &-90.8&31.0 &-90.7&54.0 \\
                 &2s &-65.6&7.86 &-65.5&28.9 &-65.4&49.9 \\
\hline
%$^{48}_\omega$Ca &1s &-110&8.25  &-110&32.6  &-110&56.9 \\ 
%                 &1p &-93.1&8.11 &-93.1&31.0 &-93.0 &53.9 \\  
%                 &2s &-69.6&7.90 &-69.6&29.2 &-69.5 &50.5 \\   
%
%\hline
$^{90}_\omega$Zr &1s &-117&8.30  &-117&33.4  &-117&58.6 \\ 
                 &1p &-105&8.19  &-105&32.3  &-105&56.5 \\  
                 &2s &-86.4&8.03 &-86.4&30.7 &-86.4&53.4 \\   
\hline
$^{208}_\omega$Pb &1s &-118&8.35 &-118&33.1 &-118&57.8 \\ 
                  &1p &-111&8.28 &-111&32.5 &-111&56.8 \\ 
                  &2s &-100&8.17 &-100&31.7 &-100&55.3 \\ 
\hline \hline
$^{6}_\omega$He &1s &-55.7&8.05 &-55.6&24.7 &-55.4&41.3 \\
\hline
$^{11}_\omega$B &1s &-80.8&8.10 &-80.8&28.8 &-80.6&49.5 \\
\hline
$^{26}_\omega$Mg &1s &-99.7&8.21 &-99.7&31.1 &-99.7&54.0 \\
                 &1p &-78.5&8.02 &-78.5&29.4 &-78.4&50.8 \\
                 &2s &-42.9&7.87 &-42.8&24.8 &-42.5&41.9 \\
\hline \hline
\end{tabular}
\end{center}
\end{table}
%
%
%%%%%%%%%%%%%%%%%%%%%%%%%%%%%%%%%%%%%%%%%%%%%%%%%%%%%%%%%%%%%%%%%%%%%%
%

Our results suggest one should expect to 
find bound $\eta$- and $\omega$-nuclear states as has been
suggested by Hayano {\it et al.}~\cite{hayano,hayano2}. 
For the $\eta$ single-particle energies, our estimated values lie 
between the results obtained using two different parameter sets 
in Ref~\cite{hayano2}. From 
the point of view of uncertainties arising from differences  
between QMC-I and QMC-II, 
the present results for both the single-particle energies 
and calculated full widths should be no more than 20 \% 
smaller in absolute values according to the estimate from the 
potential for the $\omega$ in $^{208}$Pb in Fig.~\ref{ocazrpb}.
Nevertheless, for a heavy nucleus and relatively 
wide range of the in-medium meson widths, it seems inevitable that one  
should find such $\eta$- and $\omega$-nucleus bound  states. 
Note that the correction to the real part of the single-particle 
energies from the width, $\Gamma$, can be estimated nonrelativistically, 
to be of order of $\sim \Gamma^2/8m$ 
(repulsive), which is a few MeV if we use $\Gamma \simeq 100$ MeV.

In future work we would like to include the effect 
of $\sigma$-$\omega$ mixing, which (within QHD, at least) becomes
especially important at higher densities~\cite{saitomega}.
It will also be important for consistency to calculate the in-medium
width of the meson within the QMC model and to study the energy
dependence of the meson-nucleus potential. While the energy dependence
of the potential felt (for example) by the $\omega$ may be quite
significant as we move from a virtual $\omega$ ($q^2 \sim 0$) to an
almost real $\omega$ ($q^2 \sim m^2_\omega$)~\cite{fri}, QHD studies in
nuclear matter did not reveal a strong energy dependence for $q^2$ near
$m^2_\omega$~\cite{saitomega} -- the region of 
interest here. Nevertheless, this point
merits further study in finite nuclei and within QMC itself.

To summarize, we have calculated the single-particle energies for  
$\eta$- and $\omega$-mesic  nuclei using QMC-I.
The potentials for the mesons in the nucleus have been calculated 
self-consistently in local density approximation, embedding the 
MIT bag model $\eta$ and $\omega$ mesons in the nucleus described 
by solving mean-field equations of motion. 
Although the specific form for the widths of the mesons in medium  
could not be calculated in this model yet, 
our results suggest that one should find  
$\eta$- and $\omega$-nucleus bound states for a relatively wide range 
of the in-medium meson widths.
In the near future, we plan to calculate the 
in-medium $\omega$ width self-consistently in the QMC model.
 
%
%%%%%%%%%%%%%%%%%%%%%%%%%%%%%%%%%%%%%%%%%%%%%%%%%%%%%%%%%%%%%%%%%%%%%%%%%%%%%
%
\vspace{0.5cm}

\noindent{\bf Acknowledgment}\\
We would like to thank S. Hirenzaki, H. Toki and W. Weise 
for useful discussions. Our thanks also go to R.S. Hayano for discussions 
at the {\it 2nd International Symposium on Symmetries in Subatomic Physics} 
held at University of Washington, Seattle, June 25 -- 28, 1997, which  
triggered the present work, and for providing us the experimental proposal, 
Ref.~\cite{hayano}. This work was supported by the Australian 
Research Council.
%

%%%%%%%%%%%%%%%%%%%%%%%%%%%%%%%%%%%%%%%%%%%%%%%%%%%%%%%%%%%%%%%%%%%%%%%%%%%%
%%%%%%%%%%%%%%%%%%%%%%%%%%%%%%%%%%%%%%%%%%%%%%%%%%%%%%%%%%%%%%%%%%%%%%%%%%%%
%
%\newpage
%References

%%%%%%%%%%%%%%%%%%%%%%%%%%%%%%%%%%%%%%%%%%%%%%%%%%%%%%%%%%%%%%%%%%%%%%%%%%%%%

%
\newpage

%
%%%%%%%%%%%%%%%%%%%%%%%%%%%%%%%%%%%%%%%%%%%%%%%%%%%%%%%%%%%%%%%%%%%%%%%%%%%%%
%
%%%%%%%%%%%%%%%%%%%%%%%%%%%%%%%%%%%%%%%%%%%%%%%%%%%%%%%%%%%%%%%%%%%%%%%%%%%%%
%
%
%%%%%%%%%%%%%%%%%%%%%%%%%%%%%%%%%%%%%%%%%%%%%%%%%%%%%%%%%%%%%%%%%%%%%%%%%%%%%
%
%%%%%%%%%%%%%%%%%%%%%%%%%%%%%%%%%%%%%%%%%%%%%%%%%%%%%%%%%%%%%%%%%%%%%%%%%%%%%
%
%\begin{figure}[hbt]
%\begin{center}
%\hspace*{-0.8cm}
%\epsfig{file=he6etaopot.ps,height=7cm}\,\quad
%\epsfig{file=b11etaopot.ps,height=7cm}
%\epsfig{file=mg26etaopot.ps,height=7cm}
%\caption{Potentials for the $\eta$ and $\omega$ mesons,
%($m^*_\eta(r) - m_\eta$) and ($m^*_\pomega(r) - m_\pomega$), calculated
%in QMC-I for $^{6}$He, $^{11}$B and $^{26}$Mg.}
%\label{hebmg}
%\end{center}
%\end{figure}
%
%%%%%%%%%%%%%%%%%%%%%%%%%%%%%%%%%%%%%%%%%%%%%%%%%%%%%%%%%%%%%%%%%%%%%%%%%%%%%
%
%%%%%%%%%%%%%%%%%%%%%%%%%%%%%%%%%%%%%%%%%%%%%%%%%%%%%%%%%%%%%%%%%%%%%%%%%%%%%
%
\end{document}